\documentclass[final]{svjour3}
\usepackage{graphicx}
\usepackage{rotating}
\usepackage{amssymb}
\usepackage{mathptmx}
\usepackage[numbers]{natbib}
\makeatletter
\journalname{Journal of Low Temperature Physics}

\bibpunct{}{}{,}{s}{}{,}

\begin{document}

\newcommand{\hdblarrow}{H\makebox[0.9ex][l]{$\downdownarrows$}-}
\title{Recent developments in unconventional superconductivity theory}

\author{V. P. Mineev}

\institute{Commissariat \`a l'Energie Atomique, INAC/SPSMS, \\38054
Grenoble, France\\ 
\email{vladimir.mineev@cea.fr}}

\date{08.10.2009}

\maketitle


\begin{abstract}
The review of recent developments in the unconventional superconductivity theory is given.
In the fist part I consider  
the physical origin of  the Kerr rotation polarization of light reflected from the surface of 
superconducting $\rm Sr_2RuO_4$.
Then the  comparison of magneto-optical responses in superconductors with orbital and spin spontaneous magnetization is presented.  The latter  result is applied to the estimation of the magneto-optical properties of neutral superfluids with spontaneous magnetization.
The second part is devoted to the natural optical activity or gyrotropy properties of noncentrosymmetric metals in their normal and superconducting states. The  temperature behavior of the gyrotropy coefficient
 is compared with the temperature behavior of paramagnetic susceptibility determining the noticeable increase of the paramagnetic limiting field in noncentrosymmetric superconductors.
 In the last chapter I describe the order parameter and the symmetry  
of superconducting state 
in the itinerant ferromagnet with orthorhombic symmetry.
Finally the Josephson coupling between two adjacent ferromagnet superconducting domains is discussed.

\end{abstract}

\section{Introduction}
The specific phase coherence in a macroscopic ensemble of fermions, which gives rise to non-dissipative currents in superconductors, sometimes accompanied by spacial and magnetic ordering arising after transition to the superconducting state. This type of superconductivity characterized by additional symmetry breakings is called unconventional superconductivity. The development of theory of unconventional superconductivity  has been stimulated by the experimental discoveries of superconductivity in heavy-fermion materials and  high-temperature superconductivity in layered cuprate compounds followed several years later by the discovery of superconductivity in the perovskite oxide ${\rm Sr_2RuO_4}$ with similar to cuprates structure \cite{Maeno}. Soon after, the idea 
of  the time-reversal symmetry breaking
 form of the order parameter $(\eta_x,\eta_y)\propto(1,i)$  
in this material  
  has been put forward in the paper. \cite{Sigrist}  Although there are some serious unresolved  problems related with two component superconductivity in ${\rm Sr_2RuO_4}$ (see for instance the paper \cite{Mineev08} and references therein), the idea of superconducting state with spontaneous magnetization 
 is supported by several experimental observations.  
  The most important of them are the increase of $\mu$SR zero-field relaxation rate \cite{Luke} and the Kerr rotation of reflected light from the surface of  ${\rm Sr_2RuO_4}$ in the superconducting state.\cite{Xia}.  The latter observation has stimulated intensive theoretical activity with poorly consistent  results.  In the first part  of this paper, after reviewing different theoretical approaches to this problem,  we consider the magneto-optical phenomena in superconductors with spontaneous magnetization of orbital and spin origin.

Another source of considerable interest appeared recently after the discovery of superconductivity 
 in noncentrosymmetric compounds ${\rm CePt_3Si}$ \cite{Bauer04},
${\rm UIr}$ \cite{Akazawa04}, CeRhSi$_3$ 
\cite{Kimura05}, CeIrSi$_3$ \cite{Sugitani06},
Y$_2$C$_3$ \cite{Amano04},
Li$_2$(Pd$_{1-x}$,Pt$_x$)$_3$B \cite{LiPt-PdB},
KOs$_2$O$_6$ \cite{KOsO} and others. The spin-orbit coupling of electrons in noncentrosymmetric crystal lifts the spin degeneracy of the electron energy band causing a band splitting. 
The band splitting reveals itself in the large residual value of the spin susceptibility of noncentrosymmetric superconductors at zero temperature.  This leads to the noticeable increase of paramagnetic limiting field.
 Another significant
manifestation of the band splitting is the natural optical activity or natural gyrotropy including such phenomena as double circular refraction, the Faradey and the Kerr effects.  
In the second part of this paper I consider the natural gyrotropy of noncentrosymmetric metals.
The derivation of the temperature dependence of gyrotropy coefficient in superconducting state will be given in comparison with simple and straightforward calculation of the Pauli susceptibility.

Third important direction  in the physics of unconventional superconductivity is related to the co-existence of superconductivity and itinerant ferromagnetism 
in several uranium compounds
 ${\rm UGe_2}$, \cite{Saxena} ${\rm URhGe}$, ~\cite{Aoki} and ${\rm UCoGe}$.~\cite{Huy07}. 
 It is found to arise as a co-operative phenomena rather than as the overlap of two-mutually competing orders. In the third part of the present article
 I discuss 
 the symmetry, the order parameters 
 and the interdomain Josephson coupling in the orthorhombic ferromagnet superconductor with triplet pairing.

\section{Magneto-optical properties of superconductors with spontaneous time-reversal breaking}
\subsection {Current response of the orbital origin}

The magneto-optical phenomena in a material are described in terms of   Hall conductivity, 
that is complex off-diagonal component of the conductivity tensor $\sigma_{xy}=\sigma'_{xy}+i\sigma''_{xy}$. In particular, if linearly polarized light normally  (along the $z$-direction) incident from vacuum to the boundary of a medium with complex index of refraction
 $
 N=n+i\kappa,
 $
it is reflected as elliptically polarized with the major axis rotated relative to the incident polarization by an amount  \cite{Ben}
\begin {equation}
\theta=\frac{4\pi\sigma_{xy}^{\prime\prime}}{\omega n(n^2-1)}.
\label{Kerr}
\end{equation}
This phenomenon is known as the Kerr effect. The equation for the Kerr angle is written here in the assumption $\kappa<<n$.


Magnetic superconducting states are formed by means of Cooper pairing of electrons in 
the state with nonzero orbital or spin angular momentum \cite{MinSam}. The simple example 
of the superconducting state with 
spontaneous angular momentum is so called A-phase with the order parameter
\begin{equation}
\Delta({\bf k},{\bf r})=\Delta_i({\bf r})\hat k_i, ~~~\Delta_i({\bf r})=\Delta(\hat x_i +i\hat y_i),
\label{A}
\end{equation}
 proposed for the ${\rm Sr_2RuO_4}$ superconducting state.\cite{Sigrist}
Each Cooper pair in this state possess angular momentum $\hbar \hat z$ directed along the $c$ axis of the tetragonal crystal. 
In the experiments \cite{Xia} there was studied the Kerr rotation of the reflected  linearly polarized light incident normally on the specular surface of superconducting  ${\rm Sr_2RuO_4}$ oriented parallel to the  $ab$ crystalline plane.

The calculation of the current    done in the first
papers devoted to electromagnetic properties of superconducting A-phase  \cite{Klemm88, Hir89,Hir92} in the corresponding geometry did not reveal a magneto-optical response originating from single particle excitations. The magneto-optical response
was shown to arise from the order parameter collective mode  coupled with electromagnetic field in presence of particle-hole asymmetry of energy excitation in the metallic state.\cite{Yip} The magnitude of the circular dichroism was found roughly $10^{-7}-10^{-8}~{\rm rad}$ at frequencies of the order the gap
and decreases at least as fast as $(2\Delta/\omega)^2$ at higher frequencies. This is too small in comparison with observed value\cite{Xia} of the Kerr rotation in ${\rm Sr_2RuO_4 }$ performed at frequencies about four orders of magnitude larger than the gap value.  Nevertheless the collective modes mechanism deserves more concrete investigation in application  to  ${\rm Sr_2RuO_4 }$ taking into account its quasi-two dimensional multi band  structure.

The numerous theoretical studies of the Kerr effect in ${\rm Sr_2RuO_4}$ were performed since then the experimental observations \cite{Xia} have been reported.  First, there were found the finite  Kerr angles 
\cite{Yak,Min07} of reasonable magnitude. Then the authors of more elaborate treatments \cite{Lut,Roy} 
came to the
conclusion  corresponding to 
the results of the earlier papers \cite{Klemm88, Hir89,Hir92}, namely,  that  a clean chiral $p$-wave superconductor in the spatially homogeneous in $(a,b)$ plane 
e-m field  has the vanishing  Hall conductivity and  Kerr angle. 
As result, the finite Kerr angle observed in superconducting ${\rm SrRu_2O_4}$ has been associated \cite{Lut,Roy}
with the finite size of the light spot in the experiments \cite{Xia}.

The specific magneto-optical current response to the e-m field has been found in the papers. \cite{Min08,Sig09} It originates from  the pairing term in the
 Gor'kov equations for an unconventional superconductor found in the paper  \cite{Bal}
\begin{equation}
\Delta\left({\bf k}-i\nabla-\frac{e}{c}{\bf A}({\bf r}), {\bf r}\right)F^\dagger({\bf k},{\bf r}).
\label{D}
\end{equation}
Here ${\bf r}$ is the coordinate of the Cooper pair centre of gravity, 
${\bf k}$ is the relative momentum of  two electrons in the pair. 
These two treatments \cite{Min08,Sig09} apparently return us  back to the 
finite Kerr angle found in  the earlier papers.\cite{Yak,Min07}
In fact,  the pairing term   is independent of  the vector-potential.\cite{Volovik,yak}
For the A-phase state in the coordinate representation it is
\begin{equation}
-ik_F^{-1}[\nabla_i\Delta_i({\bf r})+\Delta(_i({\bf r})\nabla_i]F^\dagger({\bf r},{\bf r}').
\end{equation}
Corresponding expression  for any unconventional superconducting state with ${\bf k}$-dependent
order parameter can be derived making use its form in ${\bf k}$ representation 
\begin{equation}
\sum_{\bf q}\Delta({\bf  k},{\bf q})F^\dagger({\bf k}-{\bf q},{\bf k}')
\end{equation}
established in \cite{MinSam}. 
Hence, the results of the papers \cite{Min08,Sig09} are incorrect.

The absence of the Hall conductivity in a superconductor with time reversal breaking of the orbital origin follows from the general arguments formulated  by Read and Green.  \cite{RG} There was pointed out that the conductivity defined as the current response to an electric field, taking the wave vector to zero before the frequency, always has the diagonal form. This result is independent of interaction whether it produces pairing or not. It can be treated 
as the contribution of the center of mass of  Cooper pair accelerated by the applied uniform electric field, while the relative motion of the particles is unaffected, as a consequence of Galilean invariance.  Hence,  the magneto-optical response in chiral superconductors state can exist only in nonuniform conditions. 
In between the treatments mentioned above there are those where the different types of inhomogeneity are essential. 

It is the case  the Hall conductivity arising
due to the order parameter "flapping" mode excitation considered by Yip and Sauls \cite{Yip}.
The flapping oscillations of the order parameter correspond to an effective inclination of the Cooper pair angular momentum from the direction perpendicular to the superconductor surface. 
This can be understood as if one edge of each 
pair proves to be closer to the metal surface than the other one. 
The e-m field attenuates in the bulk of  metal due to the skin effect.
As result the relative motion of the particles in Cooper pair is affected that yields the Hall response.

Another example of the magneto-optical response related with space inhomogeneity  was found in the authors paper \cite{Min07} where  it is given by the 
the second term in the Eqn.(23) and corresponds to the spontaneous magnetic field $H_s$
due to the textures of the order parameter arising near the boundaries of domains with up and down directions of the Cooper pairs orbital moment. The phenomenological approach making used in \cite{Min07} can be improved by calculations based on the microscopic theory taking into account  the order parameter, current and field distributions.

Also, as we already mentioned above,  the Kerr effect can be realized due to the finite size of the light spot. \cite{Lut,Roy} 

Finally, there is a possibility of the impurity induced Galilean invariance breaking.  The  polar Kerr effect  related with this mechanism has been  considered in the recent papers
by Goryo \cite{Goryo} and by Lutchyn et al \cite{Lutchyn}. 

 Which of these mechanisms gives the most important contribution to the observed Kerr rotation in the superconducting ${\rm Sr_2RuO_4}$ ?
Or, may be  we should search an another explanation of experimental observation reported in the paper\cite{Xia} ? These are at the moment the open questions.

\subsection{Current response of the spin origin}

There are also the superconducting states where the Cooper pairs have nonzero spin expectation value.
They are called nonunitary states and characterirized by complex order parameters ${\bf d}_1({\bf k})$
and ${\bf d}_2({\bf k})$ of spin up $S_z=1$ 
and spin down $S_z=-1$  states
\begin{equation}
{\bf d}_1({\bf k})=\Delta_{\uparrow}({\bf k})(\hat{x}+i\hat{y}),~~~
{\bf d}_2({\bf k})=
\Delta_{\downarrow}({\bf k})(\hat{x}-i\hat{y}).
\end{equation}
Here $\hat{x}$, $\hat{y}$  are the
unit vectors of the spin 
coordinate system.
The density of magnetic moment determined by the difference in densities of spin-up and spin-down electrons
\begin{equation}
{\bf M}_s=\mu_B(n^\uparrow-n^\downarrow)\hat z.
\end{equation}
It can be quite large if we deal with superconducting state arising from the normal ferromagnet state. This situation will be discussed in the last Chapter. Here, we assume that the magnetization spontaneously 
arises at the transtion from the normal paramagnet to the superconducting state as it is in the superfluid ${\rm A_1}$ phase of ${\rm ^3He}$ where only the pairing between spin-up ${\rm ^3He}$ atoms occurs.
Then the density of magnetic moment differs from zero  due to the assymmetry in the particle-hole distribution near the Fermi surface \cite{MinSam}
\begin{equation}
{\bf M}_s\approx\mu_BN_0^\prime\Delta^2\hat z.
\label{M}
\end{equation}
Here, $N_0^\prime$ is the derivative of electron density of states at the Fermi level.


We shall discuss the light incident normally to the surface of superconductor
with spin magnetic moment  oriented perpendicular to the surface.
To   find the Hall response due to the spontaneous spin magnetism in nonunitary superconducting state 
we need to consider the current excited by electro-magnetic field in the chargeless superfluid. 
In neglect spin-orbital coupling
 the spontaneous magnetic moment  ${\bf M}_s$ performs the free precessional motion around the  alternating transverse to its direction  magnetic field $\delta{\bf H}({\bf r},t)$. Then for the corresponding Fourier components we obtain $i\omega\delta{\bf M}=\gamma(\delta{\bf H}\times {\bf M}_s)$,
 $\delta{\bf H}=\nabla\times{\bf A}=(-ic/\omega)(\nabla\times{\bf E})$. The density of current is given by
 ${\bf j}=c(\nabla\times\delta{\bf M})$. Hence, when the electric field ${\bf E}=E(z)\hat y$, where $z$ is the direction perpendicular to the supercondactor surface, we obtain 
\begin{equation}
j_x=\frac{\gamma M_sc^2}{\omega^2}\frac{\partial^2E_y}{\partial z^2}.
\end{equation}
So, the spin part of the Hall conductivity is  
\begin{equation}
\sigma_{xy}^s\approx\frac{\gamma M_sc^2}{\omega^2\delta^2},
\end{equation} 
where $\delta$ is the skin penetration depth. Hence, as it is in the superconducting states with  an orbital time-reversal breaking, the magneto-optical response in nonunitary superconducting state arises only in the nonuniform e-m field.
Substituting the spontaneous moment density we come to the estimation 
\begin{equation}
\sigma_{xy}^s\approx \frac{e^2}{\hbar k_F\delta^2}\left ( \frac{\Delta}{\hbar\omega} \right )^2.
\end{equation} 
According to the equation (1) this result  corresponds to quite tiny Kerr rotation. 

\subsection{Magneto-optics in superfluid phases of ${\rm ^3He}$}

The results obtained in the previous subsection can be applied to the real neutral superfluids, that is to the 
superfluid ${\rm A}$ and ${\rm A_1}$ phases of ${\rm ^3He}$. Here we deal with transparent media, so we can put $\omega=cq/n$. Hence,
\begin{equation}
\sigma_{xy}^s=-\gamma n^2M_s.
\end{equation}

The Faraday polarization rotation of linearly polarized light propagating on the length $l$ along 
the  ${\bf M}_s \parallel \hat z$ direction 
is given by \cite{Ben}
\begin{equation}
\theta_F=- \frac{2\pi \sigma_{xy}^s}{nc}~l=\frac{2\pi n\gamma M_s}{c}~l.
\end{equation}
This formula is valid at $n>>\kappa$.
The Helium-3 gyromagnetic ratio is $\gamma=2\times10^4({\rm Gauss~s})^{-1}$. 
The density of  spontaneous magnetization in ${\rm ^3He-A_1}$ is  given by  eqn. (\ref{M}). The rough estimation of corresponding product $\gamma M_s$ is
\begin{equation}
\gamma M^{A_1}_s\approx 2\times10^{-2}(1-T/T_c) ~{\rm s}^{-1}.
\end{equation} 
Thus, to reach the measurable values of  the Faraday rotation due to the ${\rm ^3He-A_1}$ spontaneous magnetization  the light should run enormously long distance. 

In the superfluid ${\rm ^3He-A}$ the situation is even worst.  This state 
is an orbital ferromagnet 
where the product $\gamma M_s$ is three orders of magnitude smaller \cite{Leg, Paul} than it is in  the ${\rm ^3He-A_1}$
\begin{equation}
\gamma M_s^A\approx2\times10^{-5}(1-T/T_c)~{\rm s}^{-1}. 
\end{equation}

At the same time the product
$\gamma M_p$ corresponding to
usual paramagnetic magnetization $M_p=\chi H_0$ in the field $H_0\approx 1~{\rm Tesla}$  is about $10~ {\rm s}^{-1}$.
Hence, the measurement of the  rotation of light polarization in liquid ${\rm ^3He}$ under magnetic field is in frame of the experimental possibilities of Stanford group\cite{Xia} measured the polarization rotation 
with accuracy
of the order of 
$10^{-8}~{\rm rad}$.

\section{Noncentrosymmetric metals: susceptibility and natural optical activity}

Like the time reversal symmetry violation the space parity breaking also leads to the optical activity of medium known as  natural optical activity or natural gyrotropy.\cite{LL} Here we demonstrate this property on particular simple example of noncentrosymmetric metal with cubic symmetry in its normal and superconducting state. To introduce the basic qualities of metals without  inversion centre we begin with consideration of the Pauli susceptibility in such type of materials.

\subsection{Paramagnetic susceptibility}
The spin susceptibility in noncentrosymmetric metals in normal and superconducting states has been found by K. Samokhin .\cite{Sam07}  Here, we propose  another derivation of it free of using the field theoretical methods.
 Due to the spin-orbital coupling specific for the noncentrosymmetric crystal structure the single electron energy is the matrix
\begin{equation}
\xi_{\alpha\beta}({\bf k})=(\varepsilon({\bf k})-\mu)\delta_{\alpha\beta}+
\mbox{\boldmath$\gamma$}({\bf k})\mbox{\boldmath$\sigma$}_{\alpha\beta},
\label{matrix}
\end{equation}
 in the spin space (see for instance \cite{MinSig}). Here $\alpha,\beta=\uparrow,\downarrow$ are spin indices and $ \mbox{\boldmath$\sigma$}$ are the Pauli matrices. The pseudovector crystal field $\mbox{\boldmath$\gamma$}({\bf k})$  satisfies
$\mbox{\boldmath$\gamma$}(-{\bf k})=-\mbox{\boldmath$\gamma$}({\bf k})$ and 
$g\mbox{\boldmath$\gamma$}(g^{-1} {\bf k})=\mbox{\boldmath$\gamma$}({\bf k})$,
where $g$ is any symmetry operation of the  point group ${\cal G}$ of
the crystal. The eigenvalues of matrix (\ref{matrix})  
\begin{equation}
 \xi_{\lambda}({\bf k})=\varepsilon({\bf k})-\mu+\lambda
|\mbox{\boldmath$\gamma$}({\bf k})|, ~~~\lambda=\pm
\end{equation}
are the dispersion laws of electron spectrum splitted in two bands by the spin-orbital interaction.
The corresponding Fermi surfaces are determined by the equations $\xi_{\lambda}({\bf k})=0 $.
The difference of the band energies $2|\mbox{\boldmath$\gamma$}({\bf k}_F)|$ characterizes   the intensity of the spin-orbital coupling. The Fermi momentum taken at $\mbox{\boldmath$\gamma$}=0$ is determined by the equation $\varepsilon({\bf k}_F)=\varepsilon_F$. The electron quantum state
in each band characterizes by the spinor (eigen vector of the matrix (\ref{matrix}))
\begin{eqnarray}
\Psi_{\pm}=\frac{1}{\sqrt{2|\mbox{\boldmath$\gamma$}|(|\mbox{\boldmath$\gamma$}|-\gamma_z)}}
\left(  \begin{array}{c}\gamma_x-i\gamma_y\\ \pm|\mbox{\boldmath$\gamma$}|-\gamma_z           \end{array}\right ).
\end{eqnarray}
The spin quantization axis is  given by the unit vector $\hat {\mbox{\boldmath$\gamma$}}=\mbox{\boldmath$\gamma$}/|\mbox{\boldmath$\gamma$}|$.  
The  projections of the electron spins in two bands on the $\hat {\mbox{\boldmath$\gamma$}}$ direction have the opposite orientation
\begin{equation}
(\hat {\mbox{\boldmath$\gamma$}}({\bf k})\mbox{\boldmath$\sigma$})\Psi_{\pm}({\bf k})=\pm\Psi_{\pm}({\bf k}).
\end{equation}
In an external magnetic field the matrix of electron  energy is
\begin{equation}
\xi_{\alpha\beta}({\bf k})=(\varepsilon({\bf k})-\mu)\delta_{\alpha\beta}+
\mbox{\boldmath$\gamma$}({\bf k})\mbox{\boldmath$\sigma$}_{\alpha\beta}-{\bf h}\mbox{\boldmath$\sigma$}_{\alpha\beta}. 
\label{hmatrix}
\end{equation}
The field here is written as ${\bf h}=\mu_B{\bf H}$.
The band energies are now given by 
\begin{equation}
 \xi_{\lambda,{\bf h}}({\bf k})=\varepsilon({\bf k})-\mu+\lambda
|\mbox{\boldmath$\gamma$}({\bf k})-{\bf h}|, ~~~\lambda=\pm.
\end{equation}
Along with the changes of the band energies, the spin quantization axis is also deviated from its zero field direction
\begin{equation}
\hat {\mbox{\boldmath$\gamma$}}({\bf k})~~~\to~~~\hat {\mbox{\boldmath$\gamma$}}_{\bf h}({\bf k})=
\frac{\mbox{\boldmath$\gamma$}({\bf k})-{\bf h}}{|\mbox{\boldmath$\gamma$}({\bf k})-{\bf h}|}.
\end{equation}
The magnetic moment is written as
\begin{equation}
{\bf M}=\mu_B\sum_{\bf k}\hat {\mbox{\boldmath$\gamma$}}_{\bf h}({\bf k})\left[f( \xi_{+,{\bf h}}({\bf k}))
-f( \xi_{-,{\bf h}}({\bf k}))\right],
\end{equation}
where $f(x)$ is the Fermi distribution function.
Taking the term of the first order in magnetic field 
we obtain for the magnetic susceptibility
\begin{equation}
\chi_{ij}=-\mu_B^2\sum_{\bf k}\left\{\hat {\mbox{\boldmath$\gamma$}}_i\hat {\mbox{\boldmath$\gamma$}}_j
\left[\frac{\partial f( \xi_{+})}{\partial\varepsilon}
-\frac{\partial f( \xi_{-})}{\partial\varepsilon}\right]+(\delta_{ij}-\hat {\mbox{\boldmath$\gamma$}}_i\hat {\mbox{\boldmath$\gamma$}}_j )\frac{f( \xi_{+})
-f( \xi_{-})}{|\mbox{\boldmath$\gamma$}|}\right\}
\label{chi}
\end{equation}
The first term under the sign of summation contains the derivatives of the jumps in the Fermi distributions ${\partial f( \xi_{\pm})}/{\partial\varepsilon}=-\delta(\xi_\pm)$. The second one originates from the deviation in the spin quantization direction for the quasiparticles filling the states between the Fermi surfaces of two bands.
The explicit form  of the spin susceptibility tensor depends on vector ${\mbox{\boldmath$\gamma$}}({\bf k})$ determined by the crystal symmetry. In the simplest case of cubic symmetry
one can take ${\mbox{\boldmath$\gamma$}}({\bf k})=\gamma{\bf k}$.

Then performing summation over ${\bf k}$ for the spherical  Fermi surfaces we obtain 
\begin{equation}
\chi_{ij}=\frac{\mu_B^2}{3}\left(N_{0+}+N_{0-} -2~\frac{N_+-N_-}{|\mbox{\boldmath$\gamma$}|}                         \right)\delta_{ij}.
\end{equation}
Here $N_{0\pm}$ are the density of states at the smaller $(+)$ and the larger $(-)$ Fermi sufaces.
$N_\pm$ is the number of particles  in the corresponding band.
In the limit of small spin-orbital coupling $\gamma k_F<<\varepsilon_F$ but still at $\gamma k_F>>\mu_BH$ we come to the usual expression for the Pauli susceptibility
\begin{equation}
\chi_{ij}=2\mu_B^2N_{0}
\delta_{ij},
\end{equation}
where $N_{0}=(N_{0+}+N_{0-})/2$.

One can easily obtain the corresponding formula for the superconducting state, where the energies of quasiparticles acquire gaps $\xi_\pm \to E_\pm =\sqrt{\xi_\pm^2+\tilde\Delta_\pm^2}$.
In the simplest model with BCS pairing interaction $v_g({\bf k},{\bf k}')= -V_g$, the gap functions are the same in both bands: $\tilde\Delta_{+}({\bf k})=\tilde\Delta_{-}({\bf k})=\Delta$, the triplet component of the order parameter vanishes identically and we deal with the pure singlet pairing \cite{SamMin08} state. 
The derivatives of the Fermi distributions in the equation (\ref{chi})
yield after the integration over ${\bf k}$ the Yosida function 
$$
        Y(T)=\frac{1}{4T}\int 
        \frac{1}{\cosh^2(\sqrt{\xi_\pm^2+\Delta^2}/2T)}d\xi_\pm.
$$  
On the other hand, the energy distribution of quasiparticles far from the Fermi surfaces practically coincides with its normal state distribution. Hence, so long the band splitting is much larger than the gap $\gamma k_F>>\Delta$, the second term in the equation (\ref{chi}) written for the  
superconducting state still keeps its normal state value.  Hence, one can write for the superconducting state susceptibility
\begin{equation}
\chi_{ij}=\frac{2}{3}\mu_B^2N_{0}\left (2+Y(T)\right )
\delta_{ij}.
\end{equation}
Thus, the band splitting reveals itself in the large residual value of the spin susceptibility of noncentrosymmetric superconductors at zero temperature. As result, the  paramagnetic limiting field 
\begin{equation}
H_p=\sqrt{\frac{3}{2}}~\frac{\Delta_0}{\mu_B}
\end{equation}
is $\sqrt{3}$ times  larger than it is in the ordinary superconductors. 

Our derivation was performed for the superconducting state with pure singlet pairing. This case, the weakening of the paramagnetic suppression of superconductivity  is  connected with specific for 
the non-centrosymmetric metals
band splitting.  The simultaneous presence   of the singlet and the triplet channels in the pairing interaction leads to the formation
a superconducting state with mixed singlet-triplet pairing. Then, the paramagnetic limiting field acquires the additional increase in comparison with pure singlet pairing state.

\subsection{Natural optical activity}

 In noncentrosymmetric materials the tensor of dielectric permeability  has  linear terms in the expansion in powers of wave vector
\begin{equation}
\varepsilon_{ij}(\omega,{\bf q})=\varepsilon_{ij}(\omega,0)+i\gamma_{ijl}q_l,
\label{e1}
\end{equation}
where $\gamma_{ikl}$ is an antisymmetric third rank tensor called the tensor of gyrotropy. 
The description of the natural optical activity
in terms of linear spacial dispersion of permeability \cite{LL} is appropriate for solid or liquid dielectric media. Whereas in the case of metals, it is more natural to formulate them in terms of spacial dispersion of conductivity tensor:
\begin{equation}
\sigma_{ij}(\omega,{\bf q})=\sigma_{ij}(\omega,0)-i\lambda_{ijl}q_l.
\label{e2}
\end{equation}
The gyrotropic tensor has the most simple structure in the metals with cubic symmetry.
In this case, the usual part of the conductivity tensor is isotropic 
$\sigma_{ij}(\omega,0)=\sigma(\omega)\delta_{ij}$ and the gyrotropic conductivity tensor $\lambda_{ikl}=\lambda e_{ikl}$ is determined by the single complex coefficient  
 $\lambda=\lambda'+i\lambda''$ such that a 
normal state density of current  is 
\begin{equation}
{\bf j}={\sigma}{\bf E}+\lambda\mbox{rot}~{\bf E}.
\label{cur}
\end{equation}
The gyrotropy part of the current response to the electric field  found in \cite{MinYu} is
\begin{equation}
j^g_i(\omega,{\bf q})=2e_{ijl}e^2\gamma^2\omega
\int  \frac{d^3k}{(2\pi)^3}\hat k_l
\frac{f(\xi_+({\bf k}_+))
-f(\xi_-({\bf k}_-))}
{(\xi_+({\bf k}_+)-\xi_-({\bf k}_-))^3}
{E_j}(\omega,{\bf q}).
\label{j}
 \end{equation}
Here $f(\xi_\pm( {\bf k}_{\pm}))$ is the Fermi distribution function and $ {\bf k}_{\pm}={\bf k}\pm {\bf q}/2$.
Expanding this expression up to the first order in the components of the wave vector ${\bf q}$ and performing integration over momentum space in the limit $\hbar\omega \ll \gamma k_F\ll\varepsilon_F$, we obtain 
\begin{equation}
j^g_i(\omega,{\bf q})=e_{ijn}\frac{e^2\omega }{12\pi^2\gamma_0k_F}q_n{E_j}(\omega,{\bf q}).
\end{equation}
This corresponds to
\begin{equation}
\lambda=i\frac{e^2\omega}{12\pi^2\gamma_0k_F}.
\label{lambda}
\end{equation}

To find the Kerr rotation we shall use here the more general formula\cite{Ben} 
\begin {equation}
\theta=\frac{(1-n^2+\kappa^2)\Delta\kappa+2n\kappa\Delta n}{(1-n^2+\kappa^2)^2+(2n\kappa)^2},
\label{e3}
\end{equation} 
than that used (see eqn. (\ref{Kerr})) in the previous Section.
Here $\Delta n=n_+-n_-$ and $\Delta\kappa=\kappa_+-\kappa_-$ are the differences in the real  and imaginary parts of the refraction indices of circularly polarized lights  with the opposite 
polarization. 
For the current given by eqn. (\ref{cur}) they are
$
\Delta n=\frac{4\pi\lambda''}{c},
$
and
$
\Delta\kappa=\frac{4\pi\lambda'}{c}.
$
We see that $\Delta\kappa=0$ and  $\Delta n$
expresses through the ratio of the light frequency to the band splitting $2\gamma k_F$ as
\begin {equation}
\Delta n=\frac{\alpha}{3\pi}\frac{\hbar\omega}{\gamma k_F}.
\label{e6}
\end{equation} 
Here, $\alpha=e^2/\hbar c$ is the fine structure constant.   When the frequency of light is 
larger than the quasiparticles scattering rate (clean limit): $1<<\omega\tau<\omega_p\tau$, where $\omega_p=\sqrt{4\pi ne^2/m^*}$
is the plasma frequency,  for the Kerr angle we obtain \cite{MinYu}
\begin {equation}
\theta\approx
\frac{\alpha}{3\pi}\frac{\hbar\omega^2}{\gamma_0 k_F\omega_p^2\tau}.
\label{e75}
\end{equation}

To find the gyrotropy coefficient
in the superconducting 
state we shall follow the same procedure as was used for the spin susceptibility.
Expanding the integrand in
eqn.(\ref{j}) in powers of $\frac{\partial\xi_\pm}{\partial{\bf k}}{\bf q}$ one can note that the gyrotropic current consists of two different contributions.
One part of it is determined by the  difference of the Fermi distribution function for the quasiparticles in two bands, another one originates from the  derivatives of these functions.
The first contribution is not changed in the superconducting state, at $\Delta<<\gamma k_F$.
The second  contribution is gradually suppressed with temperature decreasing due   the gap in the superconductor quasiparticle spectrum.  As result the temperature dependence of gyrotropy coefficient in the superconducting state is
\begin{equation}
        \lambda=i\frac{e^2\omega}{8\pi^2\gamma_0k_F}\left (1-\frac{1}{3}
                Y(T)\right ),
 \label{lambdasc}                
\end{equation}
where
$Y(T)$ is the Yosida function.

\section{Ferromagnetic superconductors with triplet pairing}

The recently revealed  superconductivity  
in several uranium compounds
 ${\rm UGe_2}$, \cite{Saxena} ${\rm URhGe}$, ~\cite{Aoki} and ${\rm UCoGe}$.~\cite{Huy07} is found to arise as a co-operative phenomena rather than as the overlap of two-mutually competing orders.
 In the first two compounds the Curie temperatures $T_{Curie}$ is more than the order of magnitude higher than their critical temperatures for superconductivity. In ${\rm UCoGe}$ the ratio $T_{Curie}/T_{sc}$ at ambient pressure  is about four.
 The large exchange field and also high   upper critical field at low temperatures  strongly exceeding the 
 paramagnetic limiting field ~\cite{Huxley01,Hardy051,Huy08} 
 indicate that here we deal with Cooper pairing in the triplet state.
The triplet superconductivity in  ${\rm UGe_2}$ and ${\rm URhGe}$ coexists with itinerant ferromagnetism
such that in the pressure-temperature phase diagram the whole region occupied by the superconducting state is situated inside a more vast ferromagnetic region.
In the same family metal ${\rm UCoGe}$ the pressure dependent 
critical lines $T_{Curie}(P)$ and  $T_{sc}(P)$ of the ferromagnet and the superconducting phase transitions intersect each other. The two-band multidomain  superconducting ferromagnet state arises at temperatures below both of these lines.

The symmetries and  the 
order parameters of unconventional superconducting states
arising from the normal state with a ferromagnetic order in orthorhombic crystals with strong spin-orbital coupling have been found in the paper. \cite{Mineev} Then it was pointed out  that superconducting states in triplet ferromagnet superconductors represents a special type of two band superconducting states. \cite{Cham,Min04}.  Finally the  proper Ginzburg-Landau treatment of 
the symmetry and the order parameters of the 
paramagnet as well of the multidomain ferromagnet superconducting states was given 
\cite{Min09}. It was shown that the interband  Josephson coupling fixes the phase difference between the superconducting order parameters in two band itinerant ferromagnet. Here we 
 reconsider this phenomenon and compare it with 
the Josephson coupling between two adjacent ferromagnet superconducting domains.

\subsection{ Interband Josephson coupling}

All uranium ferromagnetic superconductors are  orthorhombic metals.
The magnetic moment in its ferromagnet state is 
directed along one crystallographic axis. We chose this direction as the $\hat z$ axis.
As it was remarked in \cite{Cham}  superconducting state in an itinerant ferromagnet represents the special type of two band superconducting state  
 consisting of pairing states formed by spin-up electrons from one band and by spin-down electrons from another band. 
Hence, a superconducting state 
characterizes by two component order parameter  
\begin{equation}
{\bf d}_1({\bf k})=\Delta_{\uparrow}({\bf k})(\hat{x}+i\hat{y}),~~~
{\bf d}_2({\bf k})=
\Delta_{\downarrow}({\bf k})(\hat{x}-i\hat{y}).
\end{equation}
Here, $\hat{x}$ and $\hat{y}$  are the
unit vectors of the spin 
coordinate system pinned to the crystal axes.

The  unconventional superconducting states
arising from the normal state with a ferromagnetic order in orthorhombic crystals with strong spin-orbital coupling belong to the two different corepresentations $A$ and $B$.\cite{Mineev}  All the states relating to the given corepresentation obey the same critical temperature. 
The order parameter amplitudes  for $A$ and $B$ states correspondingly are given by
\begin{eqnarray} 
\Delta_{\uparrow}^A({\bf k})=\eta_1(k_xu_1+ik_yu_2),\nonumber\\
 \Delta_{\downarrow}^A({\bf k})=
\eta_2(k_xu_3+ik_yu_4),
\label{e2}
\end{eqnarray}
\begin{eqnarray} 
\Delta_{\uparrow}^B({\bf k})=\eta_1(k_zv_1+ik_xk_yk_zv_2),\nonumber\\
 \Delta_{\downarrow}^B({\bf k})=
\eta_2(k_zv_3+ik_xk_yk_zv_4).
\label{e2'}
\end{eqnarray}
They are odd functions of the momentum directions of pairing particles on the Fermi
surface. The functions
$u_i=u_i(k_x^2, k_y^2, k_z^2)$ and $v_i=v_i(k_x^2, k_y^2, k_z^2)$ are invariant in respect of all transformations of orthorhombic group. We shall discuss only the $A$ state. This state  is  related to the family of nonunitary axiplanar states.

The complex order parameter amplitudes $\eta_1=|\eta_1|e^{i\varphi_1}$
and $\eta_2=|\eta_2|e^{i\varphi_2}$ are not completely independent. The  relative phase difference $\varphi_1-\varphi_2$ is chosen such that the  quadratic in the order parameter part of the Ginzburg-Landau free energy density should be minimal.  
In the case of ferromagnetic normal state the time reversal symmetry is broken and the quadratic in the order parameter components free energy density has the form 
\begin{equation} 
F=\alpha_1|\eta_1|^2+\alpha_2|\eta_2|^2+\gamma(\eta_1^*\eta_2+\eta_1\eta_2^*)
+
i\delta(\eta_1^*\eta_2-\eta_1\eta_2^*).
\label{45}
\end{equation}
Here, all the coefficients are the functions of the exchange field $h$. 
The last term breaks the time reversal symmetry.     
In the absence of exchange field $\delta=0$. 
Minimization of  free energy (\ref{45}) fixes the order parameter components phase difference $\tan(\varphi_1-\varphi_2)=\delta/\gamma$.  After substitution of this value back to (\ref{45}) we come to the expression
\begin{equation} 
F=\alpha_1|\eta_1|^2+\alpha_2|\eta_2|^2
+\sqrt{\gamma^2+\delta^2}( \eta_1^*\eta_2+\eta_1\eta_2^*).
\label{freeen}
\end{equation}
Here $\alpha_i=\alpha_{i0}(T-T_{ci})$,  $i=1,2$ are the band indices, $T_{ci}$ are the critical temperatures in each band in the absence of band mixing.
Unlike eqn. (\ref{45}) the complex amplitudes  $\eta_1=|\eta_1|e^{i\varphi}$, $\eta_2=|\eta_2|e^{i\varphi}$ in the eqn. (\ref{freeen}) have common phase factors with $\varphi=(\varphi_1+\varphi_2)/2$.
The fixation of the phase difference 
 between the band order parameters  in a superconducting 
 itinerant ferromagnet  is caused by the interband Josephson coupling. 

The free energy (\ref{freeen}) valid near the phase transition from the ferromagnet state to the ferromagnet superconducting state has been used in the papers. \cite{Cham,Min04} 
The common for the each band superconductivity critical temperature is given by
\begin{equation}
T_{sc}=\frac{T_{c1}+T_{c2}}{2}+\sqrt{\left (\frac{T_{c1}-T_{c2}}{2}\right )^2+\frac{\gamma^2+\delta^2}{\alpha_{10}\alpha_{20}}}
\end{equation}

The order parameter components  ${\bf d}_1({\bf k})$, ${\bf d}_2({\bf k})$  for the state $A$ are
invariant in respect to the following group of transformations 
\begin{equation}
G_{FS}=(E,C_2^z, RC_2^x, RC_2^y)=D_2(C_2^z),
\label{46}
\end{equation}
where
  $C_{2}^{x}, C_{2}^{y},C_{2}^{z}$ are the operations of rotation on the angle $\pi$ about the $x,y,z$- axes correspondingly.   The rotations on the angle $\pi$ about the $x$- and $y$- directions are accompanied by the time inversion $R$. 
The group of symmetry of 
superconducting ferromagnet state  is called also by  its {\it  superconducting magnetic class}. This group is the subgroup of the group of symmetry of the ferromagnet state 
$
G_F=D_2(C_2^z)\times U(1),
$
called by {\it magnetic class} \cite{LL} or the point symmetry group of the ferromagnet.  Here
$U(1)$ is the group of gauge transformations. In the superconducting state the gauge symmetry is broken.
Along with the complex conjugation the action of the 
time reversal operation $R$ on    superconducting order parameter implies also the multiplication of it by the square of its phase factor: $R\to e^{2i\varphi}R$.

Besides the state $A$, there is its time reversed state $A^*$
characterized by the complex conjugate order parameter components 
\begin{eqnarray}
{\bf d}_1^*({\bf k})=\zeta_1(\hat{x}-i\hat{y})(k_xu_1-ik_yu_2),\nonumber\\
{\bf d}_2^*({\bf k})=\zeta_2(\hat{x} +i\hat{y})(k_xu_3-ik_yu_4).
\end{eqnarray}
The states 
$A$ and  $A^*$ occupy neighboring  domains with the opposite direction of magnetization.
The state $A^*$ order parameter amplitudes are $\zeta_1=|\zeta_1| e^{i\phi_1}$ and $\zeta_2=|\zeta_2| e^{i\phi_2}$. The phase difference is fixed by  $\tan(\phi_1-\phi_2)=\delta(-h)/\gamma$. 
The superconducting states in the neighboring domains obey the same critical temperature.
The symmetry of the time reversed states $A^*$ belongs to the same 
 {\it superconducting ferromagnet class} $D_2(C_2^z)$ as the $A$-states.

\subsection{ Interdomain Josephson coupling}

Let us consider a
 flat domain wall   dividing magnetic moment up and down domains in two band ferromagnet.
 This case, the localized at $x=0$ domain wall contribution to the superconducting free energy density is given by 
\begin{eqnarray} 
F_{DW}=\left[\alpha_1(|\eta_1|^2+|\zeta_1|^2)+\gamma_1(\eta_1^*\zeta_1+\eta_1\zeta_1^*)+
i\delta_1(\eta_1^\star\zeta_1-\eta_1\zeta_1^\star)\right.\nonumber\\+
\left.\alpha_2(|\eta_2|^2+|\zeta_2|^2)+\gamma_2(\eta_2^*\zeta_2+\eta_2\zeta_2^*)+
i\delta_2(\eta_2^*\zeta_2-\eta_2\zeta_2^*)\right.\nonumber\\+
\left.\gamma_3(\eta_1^*\zeta_2+\eta_1\zeta_2^*+\eta_2^*\zeta_1+\eta_2\zeta_1^*)+
i\delta_3(\eta_1^*\zeta_2-\eta_1\zeta_2^*+\eta_2^*\zeta_1-\eta_2\zeta_1^*)\right ]\delta(x).
\end{eqnarray}
Here $\eta_{1,2}=|\eta_{1,2}|e^{i\varphi}$ 
and $\zeta_{1,2}=|\zeta_{1,2}|e^{i\phi}$ are the two bands superconducting order parameters in the left (magnetic moment-up) domain and in the right (magnetic moment-down) domain, correspondingly. In view of left-right symmetry  the modulus of the order parameters in the adjacent domains are equal 
$|\eta_{1}|=|\zeta_{1}|$ and $ |\eta_{2}|=|\zeta_{2}|$. 
The densities of the gradient energy in the left and right domains are 
\begin{equation} 
F_{grad}(x<0)=K_1\left |\frac{\partial\eta_1}{\partial x}\right |^2+
K_2\left |\frac{\partial\eta_2}{\partial x}\right |^2,
\end{equation}
\begin{equation}
F_{grad}(x>0)=K_1\left |\frac{\partial\zeta_1}{\partial x}\right |^2+
K_2\left |\frac{\partial\zeta_2}{\partial x}\right |^2.
\end{equation}
Here, the rigidity coefficients $K\sim\hbar^2/m$.
The boundary conditions at $x=0$ are derived by the minimization of the sum of domain wall and the gradient free energies
\begin{equation}
-K_1\frac{\partial\eta_1}{\partial x}=\alpha_1\eta_1+(\gamma_1+i\delta_1)\zeta_1+
(\gamma_3+i\delta_3)\zeta_2,
\end{equation}
\begin{equation}
-K_2\frac{\partial\eta_2}{\partial x}=\alpha_2\eta_2+(\gamma_2+i\delta_2)\zeta_2+
(\gamma_3+i\delta_3)\zeta_1,
\end{equation}
\begin{equation}
K_1\frac{\partial\zeta_1}{\partial x}=\alpha_1\zeta_1+(\gamma_1-i\delta_1)\eta_1+
(\gamma_3-i\delta_3)\eta_2,
\end{equation}
\begin{equation}
K_2\frac{\partial\zeta_2}{\partial x}=\alpha_2\eta_2+(\gamma_2-i\delta_2)\zeta_2+
(\gamma_3-i\delta_3)\eta_1.
\end{equation}
These boundary conditions should be substituted in the expression for the superconducting current
through the domain wall
\begin{equation}
{\bf j}=\frac{2\pi i c}{\Phi_0}\left\{K_1\left (\eta_1^*\frac{\partial\eta_1}{\partial x}-
\eta_1\frac{\partial\eta_1^*}{\partial x}\right )+(\eta_1\to\zeta_1)+
K_2\left (\eta_2^*\frac{\partial\eta_2}{\partial x}-
\eta_2\frac{\partial\eta_2^*}{\partial x}\right )+(\eta_2\to\zeta_2)\right\}.
\end{equation}
Then, after taking into account the equivalence of the pairing amplitudes $|\eta_{1}|=|\zeta_{1}|$ and$ |\eta_{2}|=|\zeta_{2}|$, we obtain
\begin{eqnarray}
{\bf j}=\frac{8\pi c}{\Phi_0}\left\{ [\gamma_1|\eta_1|^2+
\gamma_2|\eta_2|^2+\gamma_3|\eta_1||\eta_2|]
\sin(\phi-\varphi)~~~~~~~~~~~~~~~~~~~\right.\nonumber\\+\left.[\delta_1|\eta_1|^2+\delta_2|\eta_2|^2+\delta_3|\eta_1||\eta_2|]
\cos(\phi-\varphi)
\right\}
\end{eqnarray}
Thus, along with the intradomain interband Josephson coupling in the superconducting itinerant ferromagnets there is interdomain Josephson coupling responsible for the fixation of the superconducting phase difference in the neighboring domains.

 The existence of the interdomain Josephson coupling bilinear in respect of modulus of two band order parameters $|\eta_1|$ and $|\eta_2|$ is typical for the $A$ superconducting states. The order paramer for the $B$ states is vanishing in the equatorial plane $k_z=0$. This case, there is  only the higher order  Josephson coupling between the domains divided by a flat domain wall parallel to the magnetization direction.

\section{Conclusion}

In the present article we studied  the   unconventional superconducting states characterized by the time reversal breaking or by the absence of the space inversion centrum. These supeconductors  possess  the difference in the optical properties of clockwise and  counterclockwise polarized lights propagating through or reflecting from such a medium. We have discussed the optical activity 
of the superconductors with spontaneous time reversal breaking as well the natural optical activity that is the property of the superconductors with broken space parity. The latters  present 
 a specific example of multi-band superconducting state. Another particular type of multi-band superconductivity with triplet pairing also studied here is the superconductivity in the itinerant ferromagnets, where existence of the intradomain as well of the interdomain Josephson coupling has been established.
 
 \begin{acknowledgements}
I would like to thank S. A. Brazovskii, C. Kallin, K. Samokhin, V. Yakovenko,  N. Read and A. Kapitulnik  for the stimulating  information exchange.

\end{acknowledgements}

\pagebreak


\begin{thebibliography}{99.}
\bibitem{yak} Author is indebted to V. Yakovenko and C. Kallin convincingly indicated him to the erroneous form of the pairing term introduced in the paper \cite{Bal}.
\end{thebibliography}
\end{document}